# Morphology and Electronic Properties of Incipient Soot by Scanning Tunneling Microscopy and Spectroscopy

Stefano Veronesi [a, ‡], Mario Commodo [b, ‡], Luca Basta [a], Gianluigi De Falco [c], Patrizia Minutolo [b], Nikolaos Kateris [d], Hai Wang [d], Andrea D'Anna [c, *], Stefan Heun [a]

[a] NEST, Istituto Nanoscienze-CNR and Scuola Normale Superiore, Piazza San Silvestro 12, 56127 Pisa, Italy

[b] Istituto di Scienze e Tecnologie per l'Energia e la Mobilità Sostenibili, Consiglio Nazionale delle Ricerche, P.le Tecchio 80, 80125 Napoli, Italy

[c] Dipartimento di Ingegneria Chimica, dei Materiali e della Produzione Industriale, Università degli Studi di Napoli "Federico II", P.le Tecchio 80, 80125, Napoli, Italy

[d] Department of Mechanical Engineering, Stanford University, Stanford, CA94305, USA

* Corresponding author: Andrea D'Anna (anddanna@unina.it)

ABSTRACT

Soot nucleation is one of the most complex and debated steps of the soot formation process in combustion. In this work, we used scanning tunneling microscopy (STM) and spectroscopy (STS) to probe morphological and electronic properties of incipient soot particles formed right behind the flame front of a lightly sooting laminar premixed flame of ethylene and air. Particles were thermophoretically sampled on an atomically flat gold film on a mica substrate. High-resolution STM images of incipient soot particles were obtained for the first time showing the morphology of sub-5 nm incipient soot particles. High-resolution single-particle spectroscopic properties were measured confirming the semiconductor behavior of incipient soot particles with an electronic band gap ranging from 1.5 to 2 eV, consistent with earlier optical and spectroscopic observations.

1. INTRODUCTION

Formation and emission of soot from incomplete combustion of fossil and bio-derived fuels is a major concern for human health and the environment [1–5]. This has motivated a longstanding and ongoing research activity aimed at improving our understanding of the physical and chemical processes involved in soot formation. Additionally, the use of combustion-generated carbon nanoparticles for a variety of applications also generated a range of recent interests. Examples include lithium ion batteries [6] and other alternative energy systems [7]. A deeper understanding of the physicochemical properties of flame-generated carbon nanoparticles, including morphology, size, optical and electronic properties, and the relationships among them is critical to the advances in these emerging areas of research.

One of the most critical steps in soot formation is the onset of particle formation, namely inception and early growth. It is during the earlier phase of formation that flame soot attains the widest variations in terms of physicochemical properties. Immediately behind the flame front, fuel-rich flame chemistry provides the precursor species, including polycyclic aromatic hydrocarbons (PAHs), leading to the formation of the first particles [8–11]. There is a general consensus that molecular constituents of the incipient soot mainly comprise of moderately sized PAHs, as recently confirmed by high-resolution atomic force microscopy (HR-AFM) [12,13]. The size of the molecular constituents visualized by HR-AFM is in good agreement with the estimates by other experimental techniques, including high-resolution transmission electron microscopy (HRTEM) [14-16], Raman spectroscopy, UV-visible absorption spectroscopy [17-19], and mass spectrometry [20-23]. PAHs are thought to undergo chemical and physical coalescence to form clusters of larger masses [24–28]. The clustering process, or the transition from gas-phase molecules to a condensed phase, represents the nucleation process of soot particles in flames [29-32].

Considerable computational and experimental efforts have been made in order to unravel the mechanism for this critical transition. Two main nucleation mechanisms have been proposed: the chemical coalescence due to the involvement of aromatic σ and π radicals and their recombination into species, either in radical or stable form, to form cross-linked structures, and the physical coalescence due to dispersive, van der Walls interactions [29-34]. The formation of curved, fullerene-like structures, due to the inclusion of five-membered ring structures, may also contribute to the mass/molecular growth process [8,23]. Regardless of which detailed process is responsible for soot nucleation, it is largely accepted that incipient soot particles are typically a few nanometers in size. Early *in-situ* UV-visible light extinction and scattering observations [35,36] suggested the presence of particles as small as 2-3 nm during the onset of sooting process. These observations were later corroborated by measurements using on-line differential mobility analysis [37–40], synchrotron small angle X-ray scattering [41], and laser induced incandescence (LII) [42].

Measurements by transmission electron microscopy (TEM) [43], atomic force microscopy (AFM) [44-47], and helium-ion microscopy (HIM) [48,49] suggest that freshly-formed soot particles exhibit liquid-like features and contain structural inhomogeneity. These particles undergo coagulation/coalescence and surface growth leading to particle size and mass growth as their residence time in flame increases; the particle size distribution becomes bimodal [37-40] at a later time. The concurrent carbonization/dehydrogenation process that follows produces solid/mature soot particles [50-52], which aggregate into fractal-like particles eventually. The evolving nature of freshly nucleated flame soot, or nascent soot just several nanometers in size, poses significant challenges from a diagnostic point of view. TEM, and particularly high resolution TEM (HR-TEM), is effective in determining morphological and structural information for mature soot particles [14-16], but it typically fails for freshly nucleated soot particles due to ablation and annealing under high electron energy and flux [53]. Possible chemical transformations of organic materials [54] make TEM imaging of nascent soot challenging. In particular, freshly nucleated soot is expected to have a lower stability to electron beam exposure than mature soot [53]. To overcome to these limitations, Kohse-Höinghaus and coworkers [49,50] investigated nascent soot particles morphologies using helium-ion microscopy (HIM); a technique similar to the electron microscopy but efficient in achieving higher contrast and improved surface sensitivity for carbonaceous materials. HIM effectively provides information about the morphology of nascent soot based on the visualization of the particle projected-area. The results indicated that nascent soot is structurally and chemically inhomogeneous, and even the smallest particles can have shapes that deviate from a perfect sphere [49,50]. Previously, the morphologies of nascent soot have been also subject to AFM studies [44-50] (with a spatial resolution lower than that of HIM). Several computational studies [55-57] on the morphology of early soot during its coagulation/coalescence processes have also been reported.

Scanning Tunneling Microscopy (STM) is a powerful technique widely utilized for surface physics and chemistry problems. STM allows for high-resolution visualization of nascent soot without the drawback of electron-beam based techniques as stated above. A further benefit of STM is its ability to combine with Scanning Tunneling Spectroscopy (STS) for measurement of electronic band gaps at a single-particle level, thus providing useful information about the size- and morphology-dependent chemical structure of the particles investigated.

Optical band gap has been studied by UV-vis absorption spectroscopy for polydispersed nascent soot-particle films with median mobility particle sizes ranging from 2-30 nm [58,59]. Thin films of thermophoretically sampled, polydispersed nascent soot 3-20 nm in mobility diameter have been probed also by STS recently to reveal their electronic band gap [58]. Both types of measurement consistently revealed an apparent quantum confinement behavior in these particles [59]. For example, while just-nucleated soot particles have optical and electronic band-gap values in the range of 1.5 to 1.7 eV, mature soot particles assume band-gap values in the range of 0.3 to 0.5 eV [58,59]. Data interpretation from these earlier measurements was however complicated by the variations of the particle size within a sample due to the inherent size distribution of nascent soot collected from flame. This size-distribution effect impacts data interpretation, as it was shown recently that the size-dependent optical properties of the nascent soot must be modeled by considering the effect of the underlying size distribution on the absorption spectrum of a particle sample [60].

There are two specific objectives for the current work. First, high-resolution STM was utilized to visualize incipient soot particles, for the first time, on a single particle basis. Second, STS measurements were carried out to probe the electronic band gap of individual, small, single particles just a few nanometers in size and aggregates of these small particles. We sampled soot thermophoretically from an atmospheric-pressure flame onto an atomically flat substrate suitable for STS without the need for further sample preparation. The morphology of the particles was probed with some detail and compared to previous HIM observations [49,50]. The electronic band gap was obtained from the STS differential conductance spectrum for representative, individual particles, and compared to the previous measurements on polydispersed particle films.

## 2. EXPERIMENTAL METHODS

Incipient soot particles were collected from a lightly sooting laminar premixed flame of ethylene and air [12,13,18]. The flame has a cold gas velocity of 9.8 cm/s and equivalence ratio $\Phi = 2.01$, and was stabilized on a water-cooled McKenna burner. The quasi one-dimensional nature of the flame enables the temperature and species concentrations to vary only along the height above the burner, i.e., the flame residence time. Particles were collected thermophoretically by rapid insertion (insertion time 30 ms) using a pneumatic actuator; a procedure used in numerous earlier flame soot investigations [16,43-53]. Incipient soot was collected immediately behind the flame front at a distance of few millimeters from the burner surface.

The target substrate was prepared specifically for STM and STS measurements. A 20-nm thick gold film was deposited on a freshly cleaved mica substrate about 250 μm in thickness. In order to obtain an atomically flat surface, the gold-on-mica substrate is annealed under ultra-high vacuum (UHV) following the procedure described in [61]. STM and STS measurements were carried out in a RHK-VT scanning tunneling microscope (RHK Technology) at an ultra-high vacuum (UHV) base pressure of $3\times10^{-11}$ mbar. Samples were introduced in a preparation chamber at a pressure of 1× 10$^{-10}$ mbar. The preparation chamber is equipped for sample thermal annealing and connected to the STM chamber so samples can be transferred and studied without being exposed to any contaminants. In the current study, particle samples were studied initially without thermal annealing. Sample stability was verified by repeating the measurement after two months of UHV storage. Thermal annealing of the particle sample was carried out subsequently at 100 °C for 10 min in order to clean the substrate and obtain higher-contrast images.

The STM tips are prepared by electrochemical etching of a tungsten wire (diameter 250 µm), following the procedure described in [62]. The optimized set-up allows us to obtain a tip with diameter < 20 nm [63]. The as-prepared tip is degassed overnight and then flashed to remove the tungsten oxide. During the STM imaging, tip is grounded and the bias values reported in the following are applied to the sample. The STM images were collected in constant current mode. This means that the feedback moves the tip up and down to keep the tunneling current constant. Therefore, if the surface is homogeneous, the height profile corresponds to the sample topography. In case of samples deposited on a substrate, which may have different electronic properties, the profile might not accurately correspond to the topography, due to electronic effects. However, when the inhomogeneity of the electronic properties plays a relevant role, the measured height changes with bias. In this work, we did not observe strong variations with bias. The height values were within 0.05 nm over a range of bias employed. On this basis, we estimate the height uncertainty to be 0.05 nm.

To provide information about the particle ensemble to be studied by STM and STS, particle size distribution, PSD, of the incipient soot was measured on-line using horizontal high-dilution tubular probe-sampling and a differential mobility analyzer (DMA). The DMA comprises of an electrostatic aerosol classifier Vienna–type DMA (TapCon 3/150, size range of 1–40 nm), an X-ray diffusion charging source (TSI Mod. 3088), and a Faraday cup electrometer. For the selected flame condition the lowest probe-to-burner distance that could be probed without significant flame front perturbation is 7 mm from the burner surface. At that position, the PSD is unimodal, well fit by a lognormal function with the median mobility diameter $\langle d_m \rangle$ = 2 nm, and a geometric standard deviation $\sigma_g$ = 1.27, as shown in Fig. 1.

For the ex situ STM analysis, thermophoretic sampling was instead performed at 4 mm above the burner surface. Based on previous thermophoretic particle densitometry measurements [64] of the same flame, the onset of soot formation is at ~4 mm without the tubular probe. The effect of the flame perturbation by the probe and the resulting 3 mm shift has been documented and explained previously [45, 64].

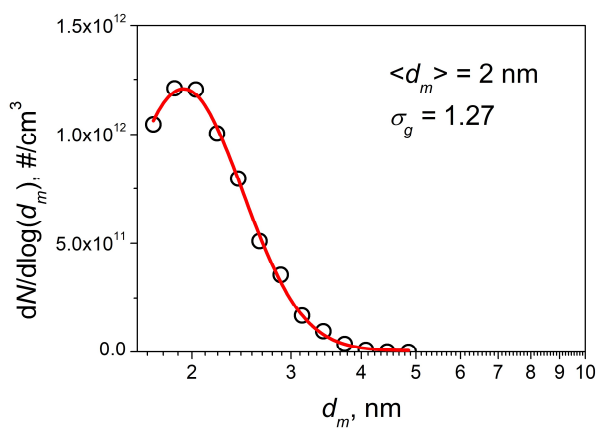

**Figure 1.** Particle size distribution of the soot sampled at the lowest burner-to-probe distance of $Z$ = 7 mm. The solid line is the lognormal fit of the data.

3.  THEORETICAL METHODS

To interpret some of the STM data, density functional theory (DFT) calculations were performed for selected PAH molecules and their clusters using the B3LYP hybrid functional [65-67] with the relativistic double-ζ effective core potential LanL2dz basis set [68], which allows the investigation of PAH interactions with gold and other transition metals. The choice for the basis set was based on considerations of a wide range of effects that needed to be explored, which include the effects of metal bonding and coordination, as will be discussed later. The LanL2dz basis set gives essentially the same results as the 6-31G(d) basis set. For example, the LanL2dz basis set yielded 3.80 eV and 4.03 eV for the HOMO-LUMO gaps of pyrene and coronene, while the 6-31G(d) basis set produced 3.85 eV and 4.04 eV, respectively. All calculations were performed using Gaussian 16 [69]. Geometry optimization was performed to obtain the molecular geometries and a frequency calculation was conducted to ensure that the optimized geometries exhibit no imaginary frequencies and lie on a minimum of the potential energy surface. Unity spin multiplicity was assumed for all species apart from free radicals in a doublet spin state due to the presence of an unpaired electron. Band gap values are calculated as the difference of the highest occupied molecular orbital (HOMO) and lowest unoccupied molecular orbital (LUMO) energies (see, e.g., [59,70,71]).

4. RESULTS AND DISCUSSION

*4.1. Size and morphological analyses*

Starting with an atomically flat substrate is critical to obtaining high-resolution images. Figure 2 shows the characterization of a typical gold-on-mica substrate. After gold-film deposition, the substrate was degassed in UHV at $1\times10^{-10}$ mbar and 120 °C overnight, followed by a ramp to 200 °C at a rate of 1 °C/min in order to achieve the Au(111)-22×√3 herringbone reconstruction. Fig. 2a shows that the surface comprises of large terraces of pristine, atomically flat gold; Fig 2b demonstrates the herringbone reconstruction observed at high resolution, consistent with previous reports [61].

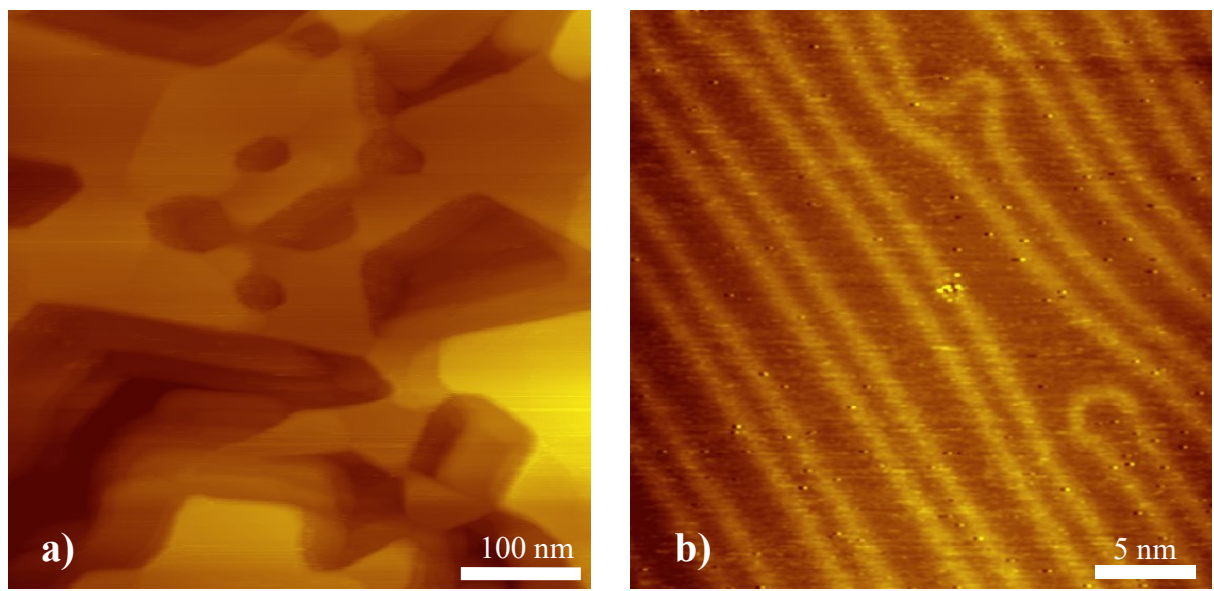

**Figure 2.** STM images acquired for a gold-on-mica substrate. a) A large-scale image (0.5×0.5 µm², image acquired using 1.3 V and 0.8 nA) showing atomically flat terraces; b) a zoom-in image (30×30 nm², 1.4 V and 0.7 nA) showing the herringbone reconstruction.

Following their thermophoretic collection, soot samples were placed in an STM sample holder and transferred to the STM preparation chamber. The samples were stored for 24 hours in UHV ($10^{-10}$ mbar) before moving to the STM stage for analysis. No significant variations in the morphology were observed for the particles over different storage times or annealing times as shown in Fig. S1 (see supplementary materials).

STM images show the presence of a variety of particles, some of which are single particles, while others exhibit aggregate features, as seen in Figs. 3a-c. The size of the particles probed here range from 2 to 8 nm, in the same range as the mobility size measurement. Even at these small sizes, the particles can exhibit features of aggregation or inhomogeneity consistent with the earlier HIM observations [48,49]. Additionally, all particles probed here are located on the terraces of the substrate. At the step edges of the gold terraces, particles often align in a row, indicating a certain affinity of the particles to the edges during sampling.

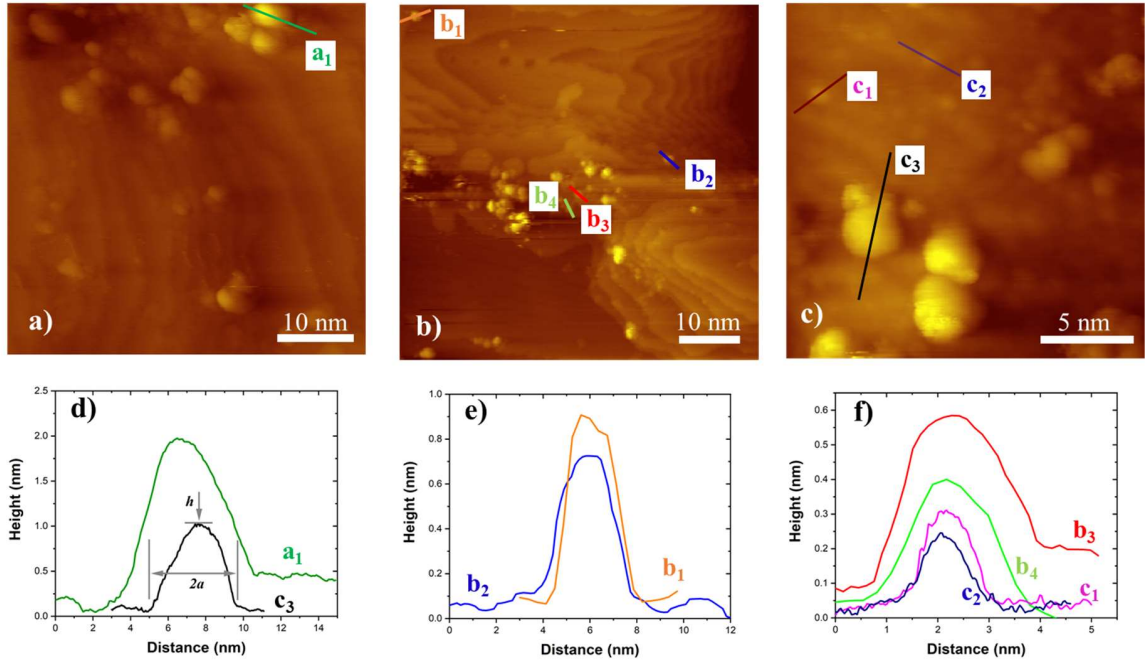

**Figure 3.** Typical STM images of incipient soot particles over the image areas of (a) 50×50 nm (image acquired using –1.6 V and –0.6 nA), (b) 60×60 nm (–1.4 V and –0.8 nA) (c) 20×20 nm (–1.8 V and –0.8 nA). Panels (d), (e) and (f) show the height profiles of various particles; the line colors correspond to those in the STM images.

The height profiles over single particles are characterized by a spherical cap with the basal diameter larger than the corresponding height, as shown in Figs. 3d and f. For all isolated particles investigated here, the morphology is rather regular. Non-spherical shapes caused by particle-particle coagulation are not as evident for particles smaller than 5 nm.

The spherical cap shape observed here is consistent with the findings from earlier studies by AFM [44-47,50]. The key reason for the observed shape is the partial flattening of a spherical particle as it strikes the substrate surface during sampling. Unlike the previous AFM studies, which show a similar morphology of surface collected particles, the finite-size tip effect is not as important for STM because of the use of extremely sharp tips and more importantly, the localized tunneling current. Assuming that the observed shape is due to a flattening of the particle material and using the topographic images acquired, the volume of the spherical cap ($v$) may be estimated from the height, $h$, and the base radius, $a$, of a spherical segment (see Fig. 3d) as $v = (\pi h/6)(3a^2+h^2)$. A volume equivalent spherical diameter $d_{eq}$ may be calculated using the measured volume [44, 45], as $d_{eq} = [h(3a^2+h^2)]^{1/3}$.

For the particles reported in Figs. 3a-c the measurement by high resolution STM results in the following volume equivalent diameter: $d_{eq} \sim 2.6$ nm for $h \sim 1.0$ nm and $a \sim 2.4$ nm (black line, panel c). The other marked particle (green line, panel a) results in a larger size with $d_{eq} \sim 4.3$ nm, ($h \sim 1.6$ nm and $a \sim 4.0$ nm). In Fig. 3b, both particles have $d_{eq} \sim 2$ nm ($h \sim 0.7$ nm and $a \sim 2.0$ nm, and $h \sim 0.8$ nm and $a \sim 1.8$ nm). These size values are consistent with those from the mobility measurement (Fig. 1). In addition to structures that are distinctively particles, the STM images also show features smaller than the particles just discussed. Typical height profiles of these features are shown in Fig. 3f. Diameters of the projected areas of these features are 1 to 1.5 nm; the height measured is consistent with that of a single atomic layer (about 0.30±0.05 nm). It is known [72,73] that the height of an atomic layer is substrate dependent because of the interaction between them. For this reason, it is not surprising to find that the PAH molecules immediately on top of the substrate have spacing smaller than the interlayer spacing of PAHs. What is accurately reproduced is the particle shape. Typically, the lateral resolution of STM is around 0.1 Å at room temperature. Hence, these smaller structures are single PAH molecules or a few of them joined in a planar cluster; they are below the mobility detection limit of 1.5 nm (Fig. 1). The size of the PAHs measured is consistent with earlier observations made with HR-AFM [12,13] in which the largest structures (e.g., IS22, IS28, IS30, IS43 and IS57) have comparable sizes.

Aggregates of several particles were also observed, as shown in Fig. 4. Given the small number of large particles observed in mobility sizing, the aggregation may have occurred during sampling when the particles were transported to the substrate, or on the substrate on which a particle undergoes surface migration upon initial impact until it strikes another particle and forms an aggregate. The primary particles in the two aggregates shown in Fig. 4a have sizes similar to those of single particles, as exemplified by the structure marked by "$S_1$" and "$S_2$" in Fig. 4a and their height profiles in Figs. 4d and 4e.

Another interesting result is that the height profile of Fig. 3c shows the characteristic, unit step height of 0.30±0.05 nm. Multiples of these unit step heights are also evident (e.g., ~0.70 nm). If the unit height is interpreted as the interlayer spacing in stacked PAHs (about 0.33 to 0.37 nm [74]), and considering a thinner first layer due to the interaction with the substrate [72,73]. The current result suggests that the STM technique has the potential to be extended to even more detailed studies of the cluster structures and sizes of incipient soot in the future.

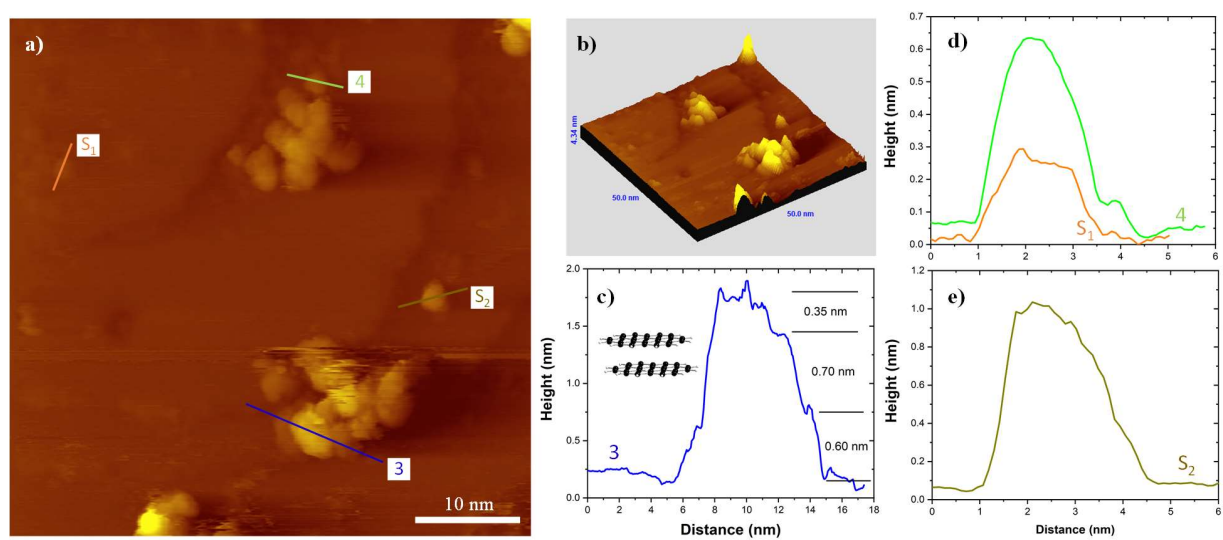

**Figure 4.** (a) STM image of representative aggregates (50×50 nm$^2$, image acquired using –1.6 V and –0.5 nA). (b) 3D reconstruction of the STM image. (c) thru (e) Height profiles along the paths marked in panel (a) with matching colors.

*4.2. Electronic properties and band gap*

Scanning tunneling spectroscopy provides a measure of single-particle, current ($I$) versus voltage ($V$) polarization curve from which the electronic band gap can be derived for individual particles. Figure 5 presents STM images, height profiles, and the corresponding STS differential conductance spectra of three representative particles. The equivalent diameters of these particles are 1.2 nm, 2.7 nm and 3.5 nm, covering an appreciable range of the particle size distribution (Fig. 1). All three particles present a spherical cap shape absent of agglomeration. For each particle, the differential conductance $dI/dV$ was directly measured by a lock-in amplifier as a function of bias $V$. The resulting spectra yield useful information about the electronic local density of states (LDOS), the edges of the conduction band (CB) and valence band (VB), and therefore the band gap $E_g$ [75]. Suffice it to note that the $E_g$ value for each particle was determined as the voltage difference between the two inflection points of the $dI/dV$-versus-$V$ curve, corresponding to the transitions from the valence band (VB) into the conduction band (CB). For details of the procedure, see [76].

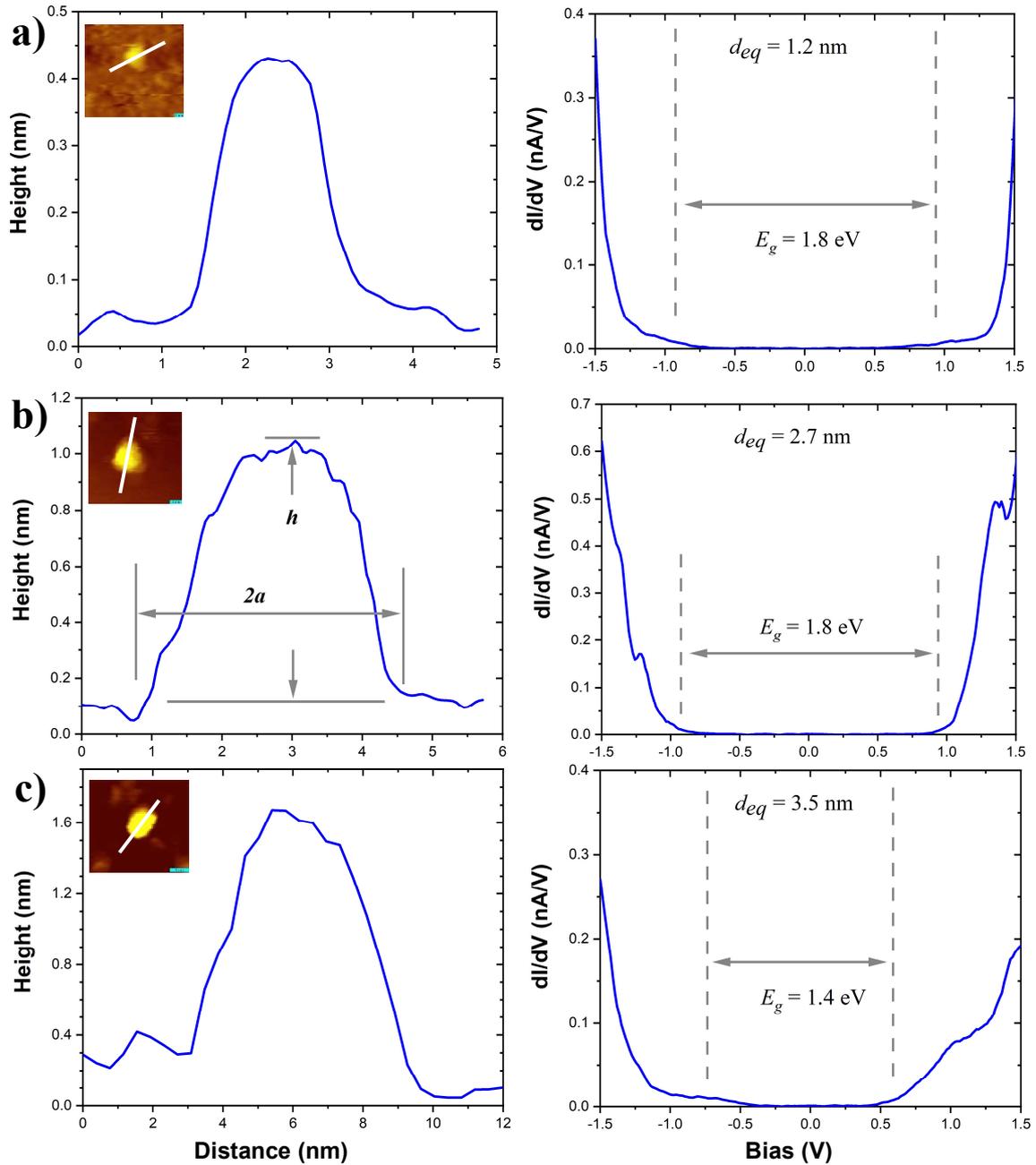

**Figure 5.** STM images (inset) and height profiles of three representative particles (left panels) and their corresponding STS differential conductance spectra (right panels). STM measurements yield size parameters of a) $2a$ = 2.5 nm and $h$ = 0.4 nm (image acquired using –1.6 V and –0.7 nA), b) $2a$ = 3.7 nm and $h$ = 0.95 nm (–1.6 V and –0.8 pA) and c) $2a$ = 6.5 nm and $h$ = 1.6 nm (–1.6 V and –0.8 nA). Also marked in the right panels are the spherical equivalent diameter of the particle probed and its corresponding electronic band gap value.

The differential conductance spectra measured here for single, isolated soot particles are consistent with those observed for the polydisperse particle film collected from the same flame [58]. Incipient particles with median diameters small than 4 nm have an electronic band gap in the range of 1.5 to 2 eV, and the band gap value decreases as the particle size increases. As indicated in each of the right panels of Fig. 5, the band gap values measured are 1.8, 1.8 and 1.4 eV, corresponding to the spherical equivalent diameter of 1.2, 2.7 and 3.5 nm, respectively. These results further support the earlier findings about the variation of the optical band gap as a function of the particle size [58,59].

Figure 6 reports the comparison of the STS spectra for several particles. It is seen that the differential conductance spectrum is asymmetric with respect to the potential bias applied for each of the particles measured. Among the particles tested, the spectra are similar in the negative potential bias region, but dissimilar for the positive bias. Suffice it to note that the negative, threshold bias at around –1 V is roughly the minimum potential value required to extract an electron from the valence band, whereas the positive threshold bias value corresponds to the energy gained from injecting an electron into the conduction band.

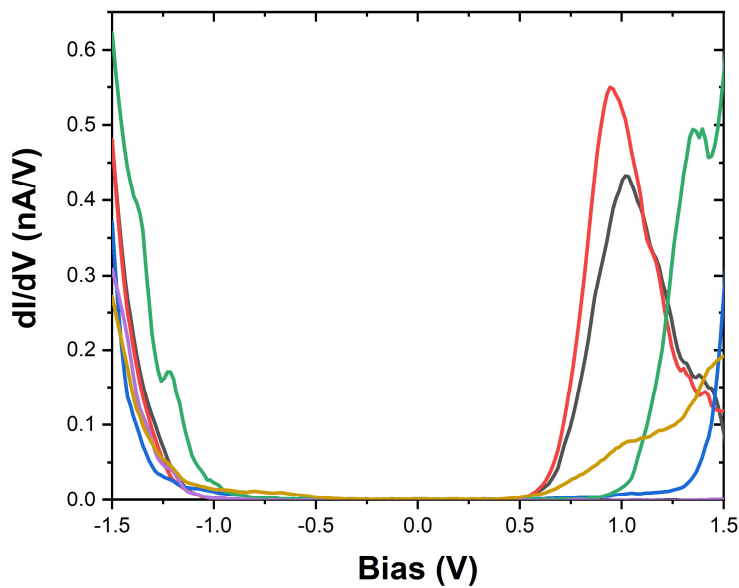

**Figure 6.** STS differential conductance spectra of several particles, all with spherical equivalent diameter < 5 nm.

The variation in the band gap observed in Fig. 6 can be attributed to a range of factors, such as a quantum confinement effect, the presence of PAHs of different sizes, and functionalized PAHs with heteroelements. The existence of an invariant valence band edge and a variable conduction band edge means that the energy required to remove an electron is the same for all particles tested, while the energy gained for electron injection differs. Hence, a plausible explanation is the variation of the electron-acceptor species, which facilitate the injection of an electron to the conduction band, while leaving the valence band unaffected. To shed light on this, DFT calculations were carried out for pyrene ($C_{16}H_{10}$) and coronene ($C_{24}H_{12}$) as the molecular model systems to investigate the above asymmetric band structure feature. The calculated values for the band gaps of pyrene and coronene are 3.80 and 4.03 eV, respectively, in good agreement with literature theoretical and experimental values [70,71,77,78]. Among potential causes examined, including the constituent PAH size and stacking effects, metal coordination, halogen substitution,

OH functionalization, and nitrogen incorporation, ketonization and free-radical were found to be consistent with the asymmetric band feature observed. As shown in Fig. 7a, ketonization lowers the LUMO energy by 0.8 eV for both pyrene and coronene, while keeping the HOMO energy constant. Figures 7b and c show the calculated density of states, obtained by applying a 0.3 eV Gaussian broadening to the molecular orbital energies of pyrene and coronene and their corresponding ketonized species. As it can be seen, the valence band does not show variation upon oxygen addition, whereas the conduction band redshifts and decreases the band gap by ~ 1 eV, a value consistent with the degree of shift in the conduction band edge in the STS spectra of Fig. 6. A similar effect is also observed for PAH $\sigma$-radicals (not shown here), in which the HOMO energy is unaffected, but the LUMO energy is reduced by 0.6 eV and 0.8 eV from those of the parent pyrene and coronene, respectively. Last but not the least, nonplanarity in PAHs can also induce a similar effect, as discussed in Xu et al. [79].

Hence, what appears to be a quantum size confinement effect could be interpreted, equivalently or in part, by the composition changes: an increased particle size increases the probability of oxygen incorporation or the presence of unpaired electrons. Furthermore, since STM/STS measurements are associated with single-electron tunneling events that are affected by the local density of states, a small concentration of oxygen or radicals can lead to a measurable change in band structure. The current interpretation is also consistent with the study of Kohse-Höinghaus and coworkers [49], in which a significant amount of surface oxygen was detected in nascent soot by X-ray photoelectron spectroscopy (XPS), including ketonic species, as seen also in other studies using different techniques (e.g., [20, 80]).

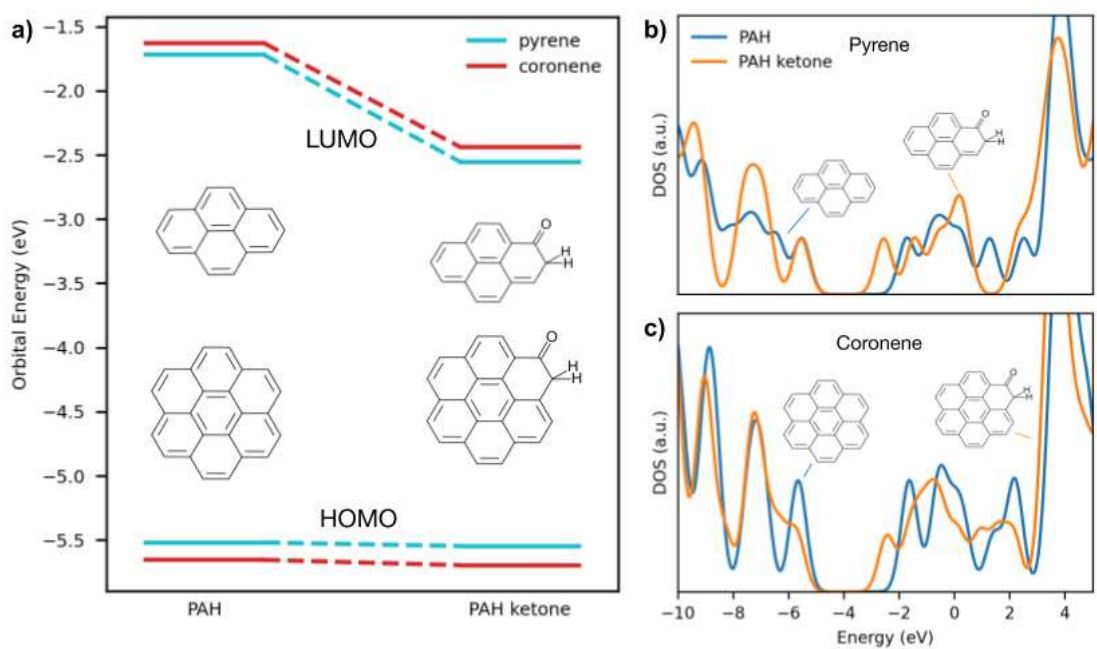

**Figure 7.** a) Shift of the HOMO and LUMO energies of pyrene and coronene upon ketonation. b) Calculated density of states of pyrene and ketonated pyrene, showing the shift in the conduction band; c) the same calculation for coronene and ketonated coronene. Zero energy corresponds to the vacuum energy as predicted by the DFT level of theory used.

## 5. CONCLUDING REMARKS

We demonstrate the use of high resolution STM for the analysis of the morphology of incipient soot in high resolution, thus extending the earlier work using other microscopic techniques [44-49]. Key conclusions are summarized in what follows:

- The size of incipient soot particles is 1 to 2 nm, in agreement with earlier estimates by in-situ optical sizing [30, 40], differential mobility analysis [31-34], AFM [40] and HIM [48,49];
- STM allows for measuring the effect of particle-substrate interactions or flattening as the particle strikes the substrate during thermophoretic sampling;
- Atomically-flat structures of 1 nm in feature size may be observed, suggesting that the STM holds the potential for probing PAH clusters formed during soot nucleation;

- Particles in the size range of just a few nanometers are rather regular in shape; non-spherical aggregates are not evident.

STS measurements were performed on single soot particles for the first time. For particles in the size range of 1 to 4 nm, the electronic band gap value was found to be in the range of 1.5 to 2 eV, and the band gap value decreases as the particle size increases. This result further supports the earlier finding about the apparent quantum confinement effect in the bandgap of flame-formed carbon nanoparticles. Further analyses of the STS results and DFT calculations suggest that the asymmetric feature of the conductance spectra and its conduction-band variation among different particles can be explained by ketonation of the constituent PAHs or by the presence of PAH $\sigma$-radicals. Hence, the apparent quantum confinement behavior observed earlier could be explained, equivalently or in part, by an increased probability of oxygenate incorporation or radical presence in the nascent soot particles.

ACKNOWLEDGMENT

The work at University Federico II in Napoli was supported by the U.S. Air Force Office of Scientific Research (AFOSR) under grant number FA8655-21-1-7022. The work at Stanford was supported by the U.S. Air Force Office of Scientific Research (AFOSR) under grant number FA9550-19-1-0261.

AUTHOR CONTRIBUTIONS

The manuscript was written through contributions of all authors. All authors have given approval to the final version of the manuscript. ‡ These authors contributed equally.